\documentclass[aps,prl,twocolumn,superscriptaddress]{revtex4-1}
\usepackage{amsmath,amssymb,amstext,graphicx,bm}

\begin{document}
\title{A phonon laser utilizing quantum-dot spin states}

\author{A.~Khaetskii}
\affiliation{Department of Physics, University at Buffalo, SUNY, Buffalo, NY 14260-1500}

\author{V.~N.~Golovach}
\affiliation{Centro de F\'{i}sica de Materiales (CFM-MPC), Centro Mixto CSIC-UPV/EHU, Manuel de Lardizabal 5, E-20018 San Sebasti\'{a}n, Spain}
\affiliation{IKERBASQUE, Basque Foundation for Science, E-48011 Bilbao, Spain}

\author{X.~Hu}
\affiliation{Department of Physics, University at Buffalo, SUNY, Buffalo, NY 14260-1500}

\author{I.~\v{Z}uti\'{c}}
\affiliation{Department of Physics, University at Buffalo, SUNY, Buffalo, NY 14260-1500}

\begin{abstract} 
We propose a nano-scale realization of a phonon laser utilizing phonon-assisted spin flips in quantum dots to amplify sound.
Owing to a long spin relaxation time, the device can be operated in a strong pumping regime,
in which the population inversion is close to its maximal value allowed under Fermi statistics.
In this regime, the threshold for stimulated emission is unaffected by spontaneous spin flips.
Considering a nanowire with quantum dots defined along its length, we show that a further improvement arises from 
confining the phonons to one dimension, and thus reducing the number of phonon modes available for spontaneous emission.
Our work calls for the development of nanowire-based, high-finesse phonon resonators.
\end{abstract} 
\maketitle

Realizing acoustic analogues of active optical devices has been a long-standing challenge.
Phonon lasers could provide versatile sources of coherent acoustic waves used for three-dimensional (3D) imaging of nanostructures
or creating periodic strain of a material to rapidly modulate its optical or electronic properties.
Recent experimental candidates include 
doped semiconductor superlattices and micro-cavity systems coupled to a radio-frequency mechanical mode~\cite{Kent,Vahala,Khurgin},
while many other possibilities have been considered theoretically~\cite{AsherScully,Zavtrak,Makler,Prieur,Chen,BargatinRoukes,ChudnovskyGaranin}.
Despite the obvious analogy between photons and (acoustic) phonons --- both being bosonic excitations with a linear dispersion,
a realization of the phonon laser is considerably more demanding.
The key difficulty stems from the small value of the speed of sound~\cite{Chen}, $s$,
or, equivalently, from the high value of the phonon density of states (DOS), 
which makes the threshold for stimulated emission hard to overcome.

A class of highly-controllable quantum systems emerging from the ideas of 
spintronics and spin-based quantum computing~\cite{DattaDas,ZuticFabianDasSarma,LossDiVincenzo,Awschalom}
may offer new regimes of physical parameters in which phonon lasing is feasible, despite the smallness of $s$.
In particular, Zeeman sub-levels of quantum dots (QDs)~\cite{HansonRMP}
have several desirable properties. 
They constitute reliable two-level systems with the spin relaxation rate 
$1/T_1$~\cite{Elzerman,Kroutvar,Hanson,Johnson,Amasha,Khaet,Khaetskii,Golovach} 
low compared to the electron tunneling rates, 
while the spin-selective tunneling~\cite{AmashaSDT,Stano,Paaske,Katsaros} 
allows to separately manipulate the populations of the spin-up and spin-down states.

The main requirements for the occurrence of stimulated emission are the following:
1) a population inversion for two levels,
2) phonon emission must dominate over other relaxation channels, 
and,
3) to overcome the threshold, the emission into the amplified phonon mode should exceed the loss due to a finite phonon lifetime, 
$\tau_Q$, in the resonator.
Usually, the latter condition is difficult to fulfill. 
Due to the high DOS for phonons,  spontaneous emission competes effectively with stimulated 
emission into the designated mode, making the population inversion small. Indeed, 
Chen and Khurgin~\cite{Chen} derive for the threshold pump rate (per unit volume),
\begin{equation}
R_{\it th}=\frac{\pi \Gamma g(\omega_Q)}{\tau_Q} \sim \frac{\omega^2_Q \Gamma}{s^3 \tau_Q},
\label{Khurgin}
\end{equation}
where $\Gamma$ is the width of the electronic level,
$g(\omega)$ is the phonon DOS in 3D, and $\omega_Q$ is the frequency of the lasing mode (throughout $\hbar=1$ and $k_{\rm B}=1$).
Remarkably, $R_{\it th}$ in Eq.~(\ref{Khurgin}) does not depend on
the interaction strength between the phonon field and the two-level system, assuming that phonon emission is the sole relaxation channel.
According to Eq.~(\ref{Khurgin}), the only realistic way of overcoming the threshold consists in using a small frequency $\omega_Q$~\cite{Chen}.
In the context of QDs discussed below, Eq.~(\ref{Khurgin}) describes the regime of {\em weak pumping}, 
which corresponds to a small value of the population inversion, and arises when the sequential-tunneling rate is small compared to $1/T_1$.

\begin{figure}[t]
\includegraphics[width=0.49\columnwidth]{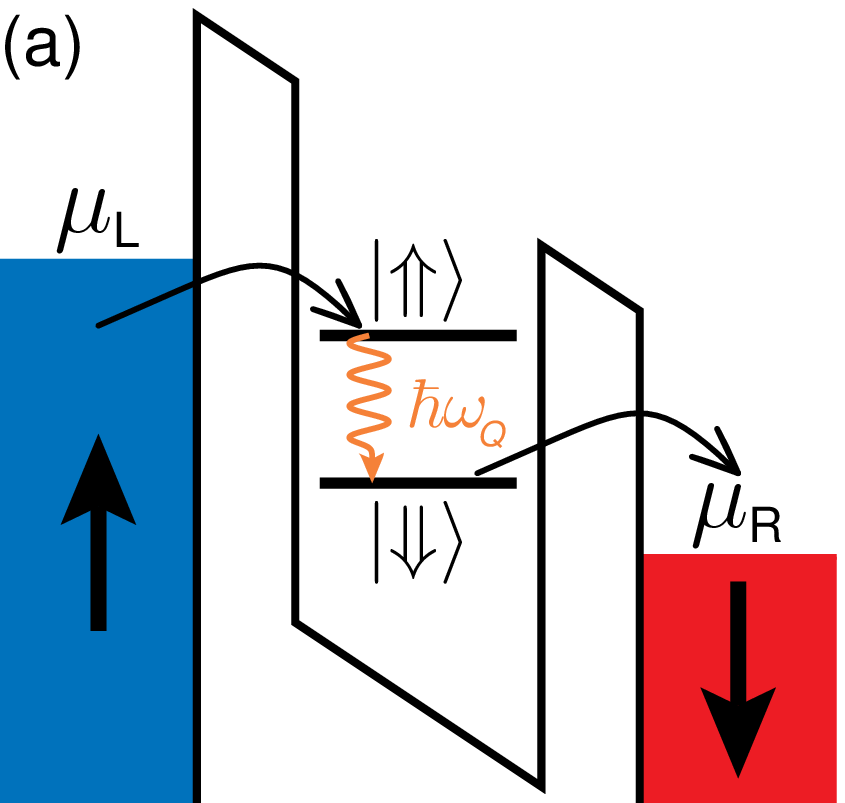}
\hfill
\includegraphics[width=0.49\columnwidth]{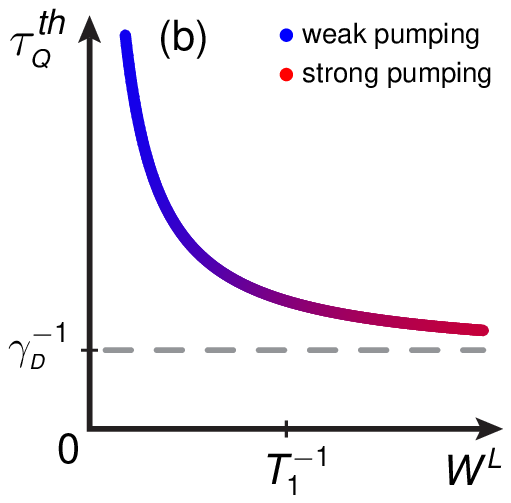}
\caption{\label{Fig1}
(a) Phonon emitter: A QD in the sequential-tunneling regime
with ferromagnetic leads of opposite polarizations and
at electro-chemical potentials $\mu_{\it L}$ and $\mu_{\it R}<\mu_{\it L}-\Delta_{\it Z}$.
The transition between the Zeeman sub-levels (from $\Uparrow$ to $\Downarrow$)
is accompanied by the emission of a phonon of frequency $\omega_{\it Q}$ matching the Zeeman splitting $\Delta_{\it Z}$.
Stimulated phonon emission dominates the transition rate in a high-finesse phonon resonator.
(b) The threshold value of the phonon lifetime $\tau_{\it Q}$ as a function of the tunnel rate $W^{\it L}$,
showing the crossover between the weak ($W^{\it L}\ll 1/T_1$) and the strong ($W^{\it L}\gg 1/T_1$) pumping regimes.
For $\tau_{\it Q}< \gamma_{\it D}^{-1}$ (below dashed line),
no lasing is possible regardless of the pumping regime.
}
\end{figure}

In this Letter we show that Zeeman sub-levels in semiconductor QDs are 
ideal two-level systems for using in phonon lasers.  
To create  population inversion, spin-selective tunneling from the leads is used [Fig.~\ref{Fig1}(a)].
Spin flip is mediated by the spin-orbit interaction in the QD
and accompanied by phonon emission~\cite{HansonRMP,Khaet,Khaetskii,Golovach}.
We find a regime of {\em strong pumping}, when the upper, spin-up,  level is occupied and the lower, spin-down, level is empty. 
This regime is accessible with QDs for realistic values of physical parameters. 
Indeed, the characteristic tunneling rates into and out of the QD that correspond to the onset of this regime are determined by small $1/T_1$ value, and can be easily adjusted. 
In this regime, the stronger the phonon field couples to the spin the lower is the threshold for stimulated emission.
Furthermore, the ability to tune the Zeeman splitting independently of the size of the QD 
allows one to control the strength of the spin-phonon coupling and 
use an {\em optimal} phonon mode for lasing.
The prescription for the angular frequency of this mode, $\omega_Q$,  is then $\omega_Q \sim s/a$, where $a$ is the QD size along the phonon propagation.

We also show that  stimulated phonon emission can be envisioned even in the weak pumping regime for sufficiently small values of the Zeeman splitting $\Delta_Z$.
However, in 3D, the threshold value in Eq.~(\ref{Khurgin}) is too demanding because the rate of spontaneous emission is too high. 
By proceeding to a 1D situation in which the phonons are emitted only along a nanowire, 
we show that $R_{\it th}$ can be strongly reduced. 
In this case all the ``wrong'' phonon modes which could have been emitted in the direction perpendicular to the propagation  of the lasing mode are excluded. 

We first consider an idealized situation depicted in Fig.~\ref{Fig1}(a), 
where a QD is tunnel coupled to two half-metal ferromagnets~\cite{note0} at electro-chemical potentials $\mu_{\it L}$ and $\mu_{\it R}$. 
The electrons from the left lead can only tunnel into the higher-energy spin-up state. 
In order to proceed to the right lead,  the electron should flip spin and transit to the lower,  spin-down, state by emitting a phonon. 
The case of ferromagnetic leads containing both spin species, and the role of the leakage current due to the spin-orbit interaction are discussed later.  
The QD Hamiltonian is
\begin{equation}
H_{\rm QD} = \sum_{s=\Uparrow,\Downarrow} \epsilon_s d_s^\dagger d_s + U n_\Uparrow n_\Downarrow,
\end{equation}
where
$\epsilon_\Uparrow = \epsilon + \Delta_Z/2$ and $\epsilon_\Downarrow = \epsilon - \Delta_Z/2$
are the energies of the two Zeeman sub-levels ($\Delta_Z>0$), 
$d_s^\dagger$ are the fermionic creation operators, and $n_s=d_s^\dagger d_s$.
The on-site Coulomb energy $U$ is assumed to be larger than the source-drain bias $eV=\mu_{\it L}-\mu_{\it R}$.
In the sequential-tunneling regime, as shown in Fig.~\ref{Fig1}(a), 
we neglect the doubly-occupied state $d_\Downarrow^\dagger d_\Uparrow^\dagger\left|0\right\rangle$,
keeping only the empty-dot $\left|0\right\rangle$,
spin-up $\left|\Uparrow\right \rangle=d_\Uparrow^\dagger\left|0\right\rangle$, and
spin-down $\left|\Downarrow\right \rangle =d_\Downarrow^\dagger\left|0\right\rangle$ states. 
The sequential-tunneling condition implies $\epsilon=0$~\cite{note1} and $eV>\Delta_Z$.

Coupling between the QD and the leads is described by the tunneling Hamiltonian
\begin{equation}
H_T = \sum_{lk\sigma s}t_{\sigma s}^l c_{lk\sigma}^\dagger d_s + \mbox{h.c.},
\end{equation}
where
$t_{\sigma s}^l$ is the matrix of tunneling amplitudes 
and $c_{lk\sigma}^\dagger$ creates an electron with momentum $k$ and spin $\sigma=\uparrow,\downarrow$ in lead $l=L,R$.
The relevant physical quantity here is the matrix of rates 
$\Gamma_{s's}^l = \pi\sum_\sigma \left(t_{\sigma s'}^l\right)^*\nu_\sigma^l t_{\sigma s}^l$,
where $\nu_\sigma^l$ is the DOS of spin species $\sigma$ in lead $l$.
In our idealized situation, only 
$\Gamma_{\Uparrow\Uparrow}^{\it L}$ and $\Gamma_{\Downarrow\Downarrow}^{\it R}$ are different from zero.
The associated sequential-tunneling rates, marked by arrows in Fig.~\ref{Fig1}(a), 
read
\begin{eqnarray}
W^L&\equiv&W_{\Uparrow 0}^L = 2\Gamma_{\Uparrow\Uparrow}^L
f(\epsilon_\Uparrow -\mu_{\it L}), \nonumber\\
W^R&\equiv&W_{0\Downarrow}^R = 2\Gamma_{\Downarrow\Downarrow}^R
\left[1- f(\epsilon_\Downarrow -\mu_{\it R})\right],
\label{rates}
\end{eqnarray}
where $f(\epsilon)=[1+\exp(\epsilon/T)]^{-1}$ is the Fermi distribution function.
The reverse rates, $W_{0 \Uparrow}^L$ and $W_{\Downarrow 0}^R$,
are obtained from Eq.~(\ref{rates}) by replacing 
$f(\epsilon) \to 1-f(\epsilon)$.
However, these rates are suppressed at low temperatures, when $T\ll eV - \Delta_{\it Z}$.

For the lasing phonon mode, we write
$H_{Q} = \omega_Q \left(N_Q + 1/2\right)$,
where $N_Q = a^\dagger a$ is the phonon number operator, 
with $a^\dagger$ creating a phonon in the lasing mode.
Well above the threshold $N_Q$ is large ($N_Q\gg 1$) and the lasing mode acts as a classical field, 
capable of driving Rabi oscillations in the QD.

The term describing the coupling between the QD and the lasing mode reads
\begin{equation}
H_a = \sum_{s's} M_{s's} d_{s'}^\dagger d_s a^\dagger + \mbox{h.c.},
\label{eqHaMssp}
\end{equation}
where $M_{s's}$ are matrix elements of the spin-phonon interaction,
obtained by taking into account a combined effect of spin-orbit interaction and magnetic field~\cite{Khaet,Khaetskii,Golovach,Trif}. 
In nanowires, such as InAs or InSb, the spin-orbit interaction is rather strong~\cite{Fasth,Nadj},
facilitating an efficient spin-phonon coupling.

The coupling of the spin to the phonon continuum, {\em i.e.}\ to all modes except the lasing mode,
is identical in nature to Eq.~(\ref{eqHaMssp}) and is obtained from Eq.~(\ref{eqHaMssp}) by summing over the phonon modes.
This coupling leads to spin relaxation~\cite{Khaet,Khaetskii,Golovach,Trif} with the rate  $1/T_1=w_{\Downarrow\Uparrow}+w_{\Uparrow\Downarrow}$,
where $w_{ss'}$ are rates for phonon-assisted transitions.
One can estimate~\cite{Khaet,Khaetskii,Golovach,Trif,note2}
\begin{equation}
w_{\Downarrow\Uparrow}\simeq 
2\pi
\left|M_{\Downarrow\Uparrow}\right|^2V g\left(\Delta_{\it Z}\right)\left[1+N(\Delta_{\it Z})\right],
\label{eqratewDnUp1pN}
\end{equation}
where $V$ is the sample volume in 3D (or length of nanowire in 1D)
and $N(\epsilon)=\left[\exp(\epsilon/T)-1\right]^{-1}$ is the Bose-Einstein distribution function.
Equation~(\ref{eqratewDnUp1pN}) gives the rate for phonon emission.
The rate for phonon absorption, $w_{\Uparrow\Downarrow}$, 
is obtained from Eq.~(\ref{eqratewDnUp1pN}) by replacing $1+N(\Delta_{\it Z})$ by $N(\Delta_{\it Z})$. 
For  low temperature, when $T\ll \Delta_Z$,  we set $w_{\Uparrow\Downarrow}=0$.  

We describe the QD by a density matrix $\hat{\rho}$, which include diagonal and off-diagonal elements.
The master equations can be derived in the standard way \cite{Blum}.  The key point is that we treat the laser mode as a classical field, assuming that its population is large $N_Q\gg 1$. 
Similar treatment for an electron coupled to an oscillating magnetic (ESR) field was used in Ref.~\onlinecite{Engel}.
After applying the rotating wave approximation, 
we obtain~\cite{RWA}
\widetext
\begin{eqnarray}
\frac{d\rho_\Uparrow}{dt} &=& W^L_{\Uparrow 0}\rho_0 - W^L_{0\Uparrow}\rho_\Uparrow + w_{\Uparrow\Downarrow}\rho_\Downarrow - w_{\Downarrow\Uparrow}\rho_\Uparrow -\gamma N_Q (\rho_{\Uparrow}-\rho_{\Downarrow}), \nonumber \\
\frac{d\rho_\Downarrow}{dt} &=& W^R_{\Downarrow 0}\rho_0 - W^R_{0\Downarrow}\rho_\Downarrow + w_{\Downarrow\Uparrow}\rho_\Uparrow - w_{\Uparrow\Downarrow}\rho_\Downarrow + \gamma N_Q(\rho_{\Uparrow}-\rho_{\Downarrow}) , \nonumber \\
\frac{d\rho_0}{dt} &=& W^L_{0\Uparrow}\rho_\Uparrow + W^R_{0\Downarrow}\rho_\Downarrow - \left(W^L_{\Uparrow 0}+ W^R_{\Downarrow 0}\right)\rho_0, \nonumber \\
\frac{dN_Q}{dt} &=&   \gamma N_Q (\rho_{\Uparrow}-\rho_{\Downarrow})  -\frac{N_Q}{\tau_Q}. 
\label{system}
\end{eqnarray}
\endwidetext
\noindent
where $\rho_\Uparrow + \rho_\Downarrow +\rho_0 = 1$ is due to Coulomb blockade and 
$\gamma=2|M_{\Downarrow\Uparrow}|^2 /\Gamma$. 
The quantity $\gamma N_Q$ is the rate of Rabi flips.
The quantity $\Gamma= (W^L_{0\Uparrow} +W^R_{0\Downarrow})/2 + 1/T_2$ 
is the decay rate of the off-diagonal component $\rho_{\Uparrow\Downarrow}$ of the density matrix.  
It includes the component due to tunneling from the up and down spin states to the left and right leads, see Eq.~(\ref{rates}),  
and intrinsic decoherence rate $1/T_2$.
The last equation of system (\ref{system}) describes the occupation of the lasing mode. 
The decay rate $1/\tau_Q$ represents the loss of phonons due to scattering processes, including  escape through the mirrors. 

Equation~(\ref{system}) is written for  a single QD in the system. 
When there are $N_D$ identical QDs, and distance between them is larger than the phonon wave length, 
then in the system of equations for quantities $\hat{\rho}$  and $N_Q/N_D$,  $\gamma$ is replaced everywhere by $\gamma_D=\gamma N_D$. 
Since $\gamma$ is proportional to the coupling constant $|M_{\Downarrow\Uparrow}|^2$, 
it means that the normalization volume (for the phonon wave function) which enters the problem is equal to $n^{-1}=V/N_D$, i.e. the volume per one QD. 

Next, we seek a stationary solution of Eq.~(\ref{system}) and conditions for the onset of stimulated emission.
We do not present the explicit expressions for $\hat{\rho}$  and give  only the equation for the population inversion.  
For non-trivial solutions, for which $N_Q$ does not vanish identically, 
from the last line in Eq.~(\ref{system}) we obtain
\begin{equation}
\rho_\Uparrow - \rho_\Downarrow = 1/(\gamma_D\tau_Q). 
\label{EqrhoUpmrhoDn}
\end{equation}
This equation has a simple physical meaning, namely, in the stationary regime the incoming rate to the lasing mode should be equal to the decay rate $1/\tau_Q$.   
The number of phonons per QD reads
\begin{equation}
N_Q/N_D=\frac{\left(W^R-1/T_1\right)\left(\tau_Q-1/\gamma_D\right)}{2+W^R/W^L}-\frac{1}{\gamma_D T_1}.
\end{equation}
From the condition $N_Q>0$ one gets
\begin{equation}
\frac{1}{\gamma_D\tau_Q} <\frac{T_1W^R-1}{T_1W^R+1+W^R/W^L},
\label{threshold}
\end{equation}
determining the threshold value of $\tau_Q$ that corresponds to the onset of the stimulated phonon emission [see Fig.~\ref{Fig1}~(b)]. 
The quantity $T_1W^R$ should be larger than unity.  
Further we assume the inequality $T_1W^R\gg 1$.
\par 
Two pumping regimes can be distinguished [see Fig.~\ref{Fig1}~(b)].  The weak pumping  $T_1W^L\ll 1$ corresponds to almost empty dot, $\rho_0 \approx 1$, and Coulomb blockade does not play any role. 
From Eq.~(\ref{threshold})  the threshold value of the pump rate  $W^L_{\rm th}$ is 
\begin{equation}
W_{\rm th}^L = \frac{1}{T_1}\frac{1}{\gamma_D\tau_Q} \sim \frac{g \Gamma}{\tau_Q n}, 
\label{threshold3}
\end{equation}
 where we used $1/T_1 \sim |M_{\Downarrow\Uparrow}|^2 g V$ and  $g$ is the phonon DOS calculated at the Zeeman energy. 
Eq.~(\ref{threshold3}) is valid for the case of any dimensionality, $n$ is the corresponding concentration of the QDs. 
  Note that Eq.~(\ref{threshold3}) is similar to Eq.~(\ref{Khurgin}), and the coupling constant drops out. 

The condition Eq.~(\ref{threshold3}) has a simple physical meaning, and can be derived in the following way. 
At the threshold, the incoming rate to the lasing mode should exceed the decay rate $\rho_{\Uparrow}\gamma_D >1/\tau_Q$. 
On the other hand, the pump rate at the threshold is equal to the spontaneous emission rate, $W^L_{\rm th}=\rho_{\Uparrow}/T_1$.   
Excluding $\rho_{\Uparrow}$ from these equations, one obtains Eq.~(\ref{threshold3}).
 
\begin{figure}
\includegraphics[width=0.6\columnwidth]{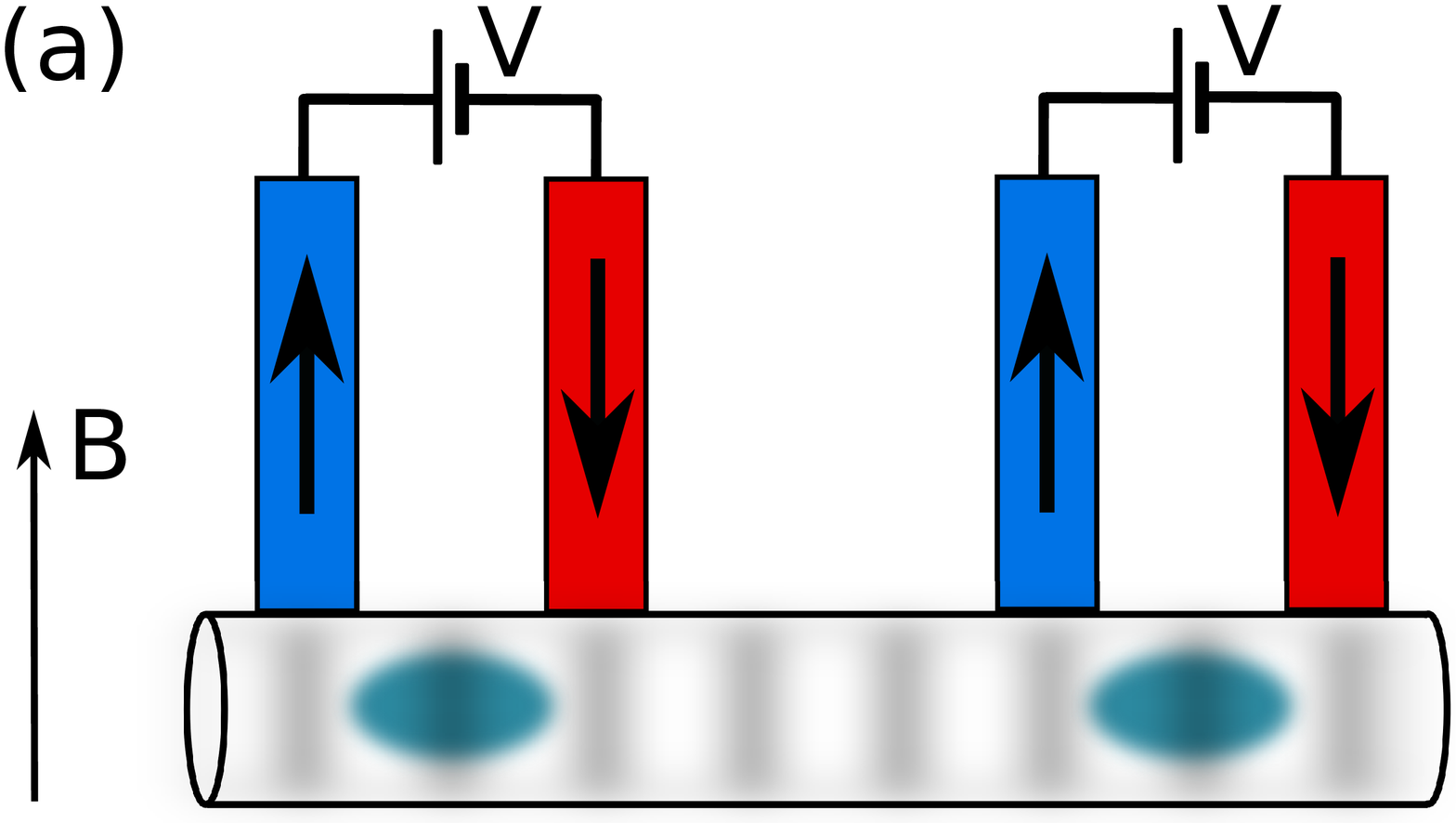}
\includegraphics[width=0.4\columnwidth]{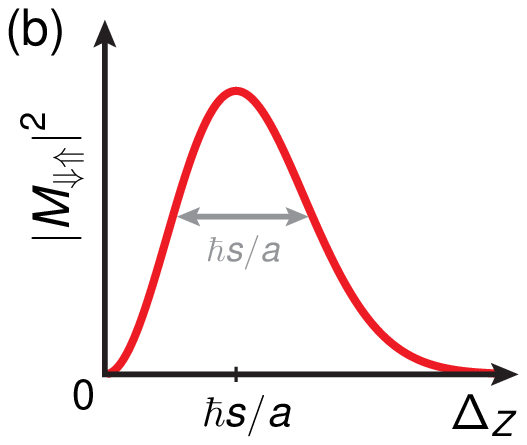}
\caption{\label{Fig2}
(a) Phonon nano-laser: QDs defined along a
nanowire and contacted by ferromagnetic fingers. Phonons
are emitted along the nanowire and are reflected at its ends.
Stimulated emission begins when the phonon loss in the
nanowire is reduced below a threshold value.
(b) Coupling constant $\left|M_{\Downarrow\Uparrow}\right|^2$ versus the Zeeman splitting $\Delta_{\it Z}$;
maximal coupling is achieved at $\Delta_{\it Z}\simeq \hbar s/a$.
}
\end{figure}

For the ratio of the threshold pump rates in 3D and 1D we obtain
\begin{equation}
\frac{W_{\rm th, 3D}^L}{W_{\rm th, 1D}^L}  \sim \frac{g_3 n_1}{g_1 n_3}\sim \frac{A}{\lambda_{ph}^2} \gg 1, 
\label{ratio}
\end{equation}
where  $g_1 =1/\pi s$ is the 1D phonon DOS, 
$\lambda_{ph}$ is the phonon wave length,  
and $A$ is the area of the sample in the transverse direction (perpendicular to the direction of the lasing mode propagation between the mirrors). 
The ratio in Eq.~(\ref{ratio}) represents the number of phonon modes which are emitted in the transverse direction and are useless for the lasing regime. 
Therefore, one can greatly reduce the threshold pump value by proceeding to a situation when the phonons propagate only in the relevant direction, such as  along a nanowire [see Fig.~\ref{Fig2}(a)]. 

In the opposite regime of strong pumping $T_1W^L\gg 1$,  the effective {\it threshold} pump rate $W_{th}^L \rho_0$ of the upper level saturates at the $1/T_1$ value.   
In this regime $\rho_\Uparrow \approx 1$, and because of the Pauli principle an electron cannot tunnel from the left lead into the QD until the electron inside the QD flips its spin and gets to the lower level, leaving the QD. 
Thus, spin-flip transition, which happens with the rate $1/T_1$,   is the bottle-neck process in this regime.  From the condition that the  incoming rate to the lasing mode exceeds the decay rate, we obtain for  the  threshold value of $\tau_Q$  the following inequality: $1/\gamma_D \tau_Q^{\rm th}<1$. 
This inequality follows also from Eq.~(\ref{threshold}).  
Note that the phonon DOS drops out, and the condition for $\tau_Q^{\rm th}$ is determined only by the coupling constant $|M_{\Downarrow\Uparrow}|^2$, 
shown as function of $\Delta_Z$ in Fig.  \ref{Fig2}(b). 
The dependence of threshold value of $\tau_Q$ on the tunneling rate $W^L$ is shown in Fig.~\ref{Fig1}(b). 
The plateau value of $\tau_Q^{th}\propto 1/\gamma_D$  can be reduced by choosing the material with a strong spin-orbit interaction like InAs. 
The curve corresponding to a much higher value of $1/T_1$ and a small value of population inversion is described by  Eq.~(\ref{Khurgin}), and  goes above the plateau in the strong tunneling regime in Fig.1(b).

For strong pumping the threshold condition can be rewritten  in 
terms of the phonon mean free path ($l_{ph}^{\rm th}=s \tau_Q^{\rm th}$ ). Specifically, for 1D, 
\begin{equation} 
N_D\frac{l_{ph}^{\rm th}}{L_z}=\Gamma T_1, 
\label{thresholdQB}
\end{equation}
where $L_z$ is the distance between the mirrors.  
 Assuming also that $ \gamma_D \tau_Q \gg 1$, i.e. well above the threshold,  we obtain for the number of phonons  in  strong tunneling regime
\begin{equation}
N_Q \approx \frac{W^L W^R}{2W^L +W^R} \tau_Q N_D.
\label{number}
\end{equation}

Our consideration so far did not take into account the leakage  current, i.e. when the spin-up electron can tunnel  directly to the right lead   without flipping its spin inside the QD,  
and an electron from the left lead can directly tunnel into the spin down state of the QD. 
Such processes are possible for minority carriers in ferromagnets and because with the  spin-orbit interaction the spin-up/-down directions are not exactly collinear. 
Adding the corresponding terms into system Eq.~(\ref{system}), with the tunneling rates $\tilde{W}_{\Downarrow 0}^L \equiv \tilde{W}^L$ 
and  $\tilde{W}_{ 0 \Uparrow}^R \equiv \tilde{W}^R$, we derive a new threshold equation instead of Eq.~(\ref{threshold}). 
In the case of relatively strong leakage when $ \tilde{W}^R \gg 1/T_1$, the threshold equation takes the form
\begin{equation}
\frac{1}{\gamma_D\tau_Q } <\frac{W^LW^R-\tilde{W}^L \tilde{W}^R}{W^LW^R+\tilde{W}^L \tilde{W}^R+ W^R \tilde{W}^R}.
\label{leakage}
\end{equation}
We see that even in the case of strong spin-orbit coupling, 
when $\tilde{W}^R $ and $W^R$ are of the same order of magnitude, 
but not very close to each other ($\tilde{W}^R < W^R$, ~$\tilde{W}^L <W^L$), 
the threshold condition is similar to that we had before in the strong tunneling case without leakage, 
when the right hand side of Eq.~(\ref{leakage}) was unity. 

Finally, the number of  nonradiative (nrd) and radiative (rad) transitions per unit time inside the QD are 
$I_{\rm nrd}= w_{\Downarrow\Uparrow}\rho_\Uparrow-w_{\Uparrow\Downarrow}\rho_{\Downarrow};\,\,
I_{\rm rad}= \gamma N_Q\left(\rho_\Uparrow-\rho_{\Downarrow}\right) \equiv N_Q/(N_D\tau_Q)$. 
A figure of merit of the emitter is the ratio $\eta=I_{\rm rad}/I_{\rm nrd}$. In the most favorable case this ratio 
reaches $\eta\approx (2T_1/\tau_Q)(N_Q/N_D)\approx T_1W^R$. 
The power output of the phonon laser is written as
\begin{equation}
P=\hbar\omega_Q N_Q/\tau_Q.
\label{power}
\end{equation}

We next  estimate the relevant parameters. 
Since the typical length of the QDs is $a \approx (0.3 \div 1)\times 10^{-5} cm$, in order to have reasonably strong coupling to the phonons one needs to choose not very large Zeeman gap $\Delta_Z \simeq \hbar s/a$, ~ $1K <\Delta_Z <5K$ [see Fig.~\ref{Fig2}(b)].   Therefore, the temperature is also restricted to these values. To have not very high value of the threshold phonon mean free path, we take the tunneling rate $\Gamma =(10^9\div 10^{10})s^{-1}$, which corresponds to the current $I=e\Gamma \approx 1 nA$. 
Then, assuming for InAs QDs relatively short $T_1\simeq (10^{-7}\div 10^{-8}) s$, and taking $N_D=10$,  
we obtain from Eq.~(\ref{thresholdQB}) the ratio $l_{ph}^{\rm th}/L_z =10$.  
To find how realistic is that, we take $L_z=1\mu m$, and $\tau_Q\simeq 10^{-7} s$ \cite{mirrors}, which corresponds to a phonon mean free path $(10^{-2}\div 10^{-1}) cm$. Then we obtain $l_{ph}/L_z \sim (10^2\div 10^3)$.    The indicated values for $l_{ph}$ were experimentally observed \cite{Kent1,Zeitler,Kittel} for  
the THz acoustic phonons in 3D semiconductors.  
For the number of phonons  above the threshold, 
see Eq.~(\ref{number}), we get $N_Q\simeq N_D \Gamma \tau_Q \simeq 10^3$. 
For the power, see Eq.~(\ref{power}), one gets $P\approx 4.2 \times 10^{-6} erg/sec$ for $\Delta_Z =3K$. 
Taking the diameter of a wire $10^{-6}cm$, we obtain for the power density $\approx 1W/cm^2$.
\par

In conclusion, we propose to use the Zeeman sub-levels of the ground orbital state of the QD 
to generate stimulated phonon emission.
The frequency of phonons can be easily tuned by changing the external Zeeman field, 
which allows  a reasonably large interaction with phonons for a given QD size [see Fig.~\ref{Fig2}(b)].    
Because of a generally low value of spin-relaxation rate,  
the strong pumping regime, characterized by a large value of the population inversion,  
can be easily achieved. 
We show that a promising practical implementation is a system of elongated QDs embedded into 1D nanowire. 
The threshold for stimulated emission is greatly  reduced in the 1D case, when the phonons propagate only along the wire. 

A.~Kh. acknowledges the financial support from  SPINMET project (FP7-PEOPLE-2009-IRSES).  
The work was supported by ONR, AFOSR, DOE-BES, and NSF-ECCS. X.~Hu also acknowledges support by US ARO and NSF-PIF.


\begin{thebibliography}{99}

\bibitem{Kent}
R. P.~Beardsley, A. V.~Akimov, M.~Henini, A. J. ~Kent,   Phys.\ Rev.\ Lett.\ \textbf{104}, 085501 (2010).

\bibitem{Vahala}
I. Grudinin, H. Lee, O. Painter, K. J. Vahala,  Phys.\ Rev.\ Lett.\ \textbf{104}, 083901 (2010).

\bibitem{Khurgin}
J. B.~Khurgin,  Physics \textbf{3}, 16 (2010). 

\bibitem{AsherScully}
I.M.~Asher and M.O.~Scully,
Phys.\ Rev.\ A \textbf{8}, 1988 (1973).

\bibitem{Zavtrak}
S.T.~Zavtrak, 
Phys.\ Rev.\ E \textbf{51}, 2480 (1995).

\bibitem{Makler} 
S.S.~Makler, D.E.~Tuyarot, E.V.~Anda, and M.I.~Vasilevskiy, 
Surf. Sci., \textbf{361/362}, 239 (1996).

\bibitem{Prieur}
J.Y.~Prieur, M.~Devaud, J.~Joffrin, C.~Barre, M.~Stenger, and M.~Chapellier,
Physica B, \textbf{219{\&}220}, 235 (1996).

\bibitem{Chen}
J.~Chen and J. B.~Khurgin, IEEE J.\ Quantum Elect.\ \textbf{39}, 600 (2003). 

\bibitem{BargatinRoukes}
I.~Bargatin and M.L.~Roukes,
Phys.\ Rev.\ Lett.\ \textbf{91}, 138302 (2003).

\bibitem{ChudnovskyGaranin}
E.M.~Chudnovsky and D.A.~Garanin,
Phys.\ Rev.\ Lett.\ \textbf{93}, 257205 (2004).

\bibitem{DattaDas}
S.~Datta and B.~Das,
Appl.\ Phys.\ Lett.\ \textbf{56}, 665 (1990).

\bibitem{ZuticFabianDasSarma}
I.~\v{Z}uti\'{c}, J.~Fabian, and S.~Das Sarma,
Rev.\ Mod.\ Phys.\ \textbf{76}, 323 (2004).

\bibitem{LossDiVincenzo}
D.~Loss and D.P.~DiVincenzo,
Phys.\ Rev.\ A \textbf{57}, 120 (1998).

\bibitem{Awschalom}
D.D.~Awschalom, D.~Loss, N.~Samarth, eds., 
{\it Semiconductor Spintronics and Quantum Computing} 
(Springer, New York, 2002).

\bibitem{HansonRMP}
R.~Hanson, L.P.~Kouwenhoven, J.R.~Petta, S.~Tarucha, and L.M.K.~Vandersypen,
Rev.\ Mod.\ Phys.\ \textbf{79}, 1217 (2007).

\bibitem{Elzerman}
J.M.~Elzerman, R.~Hanson, L.H.~Willems van Beveren, 
B.~Witkamp, L.M.K.~Vandersypen, and L.P.~Kouwenhoven,
Nature \textbf{430}, 431 (2004).

\bibitem{Kroutvar}
M.~Kroutvar, Y.~Ducommun, D.~Heiss, M.~Bichler, D.~Schuh,
G.~Abstreiter, and J.J.~Finley, 
Nature \textbf{432}, 81 (2004).

\bibitem{Hanson}
R.~Hanson, L.H.~Willems van Beveren, 
I.T.~Vink, J.M.~Elzerman, W.J.M.~Naber, F.H.L.~Koppens, L.P.~Kouwenhoven, and L.M.K.~Vandersypen,
Phys.\ Rev.\ Lett.\ \textbf{94}, 196802 (2005).

\bibitem{Johnson}
A.C.~Johnson, J.R.~Petta, J.M.~Taylor, A.~Yacoby, M.D.~Lukin, C.M.~Marcus, M.P.~Hanson, and A.C.~Gossard,
Nature \textbf{435}, 925 (2005).

\bibitem{Amasha}
S.~Amasha, K.~MacLean, I.P.~Radu, D.M.~Zumb\"{u}hl, M.A.~Kastner, M.P.~Hanson, and A.C.~Gossard,
Phys.\ Rev.\ Lett.\ \textbf{100}, 046803 (2008).

\bibitem{Khaet}
A.V.~Khaetskii and Yu.V.~ Nazarov,    Phys.\ Rev.\ B \textbf{61}, 12639    (2000). 

\bibitem{Khaetskii}
A.~Khaetskii and Yu.~ Nazarov,    Phys.\ Rev.\ B \textbf{64}, 125316     (2001). 

\bibitem{Golovach} 
V.N.~Golovach, A.~Khaetskii, and D.~Loss, 
Phys.\ Rev.\ Lett.\ \textbf{93}, 016601 (2004).

\bibitem{AmashaSDT}
S.~Amasha, K.~MacLean, I.P.~Radu, D.M.~Zumb\"{u}hl, M.A.~Kastner, M.P.~Hanson, and A.C.~Gossard,
Phys.\ Rev.\ B \textbf{78}, 041306 (2008).

\bibitem{Stano}
P.~Stano and P.~Jacquod,
Phys.\ Rev.\ B \textbf{82}, 125309 (2010).

\bibitem{Paaske}
J.~Paaske, A.~Andersen, and K.~Flensberg,
Phys.\ Rev.\ B \textbf{82}, 081309 (2010).

\bibitem{Katsaros}
G.~Katsaros, V.N.~Golovach, P.~Spathis, N.~Ares, M.~Stoffel, F.~Fournel, O.G.~Schmidt, L.I.~Glazman, and S.~De Franceschi,
Phys.\ Rev.\ Lett.\ \textbf{107}, 246601 (2011).

\bibitem{note0}
In practice, normal-metal leads can also be used, since quantum dots with spin-orbit interaction exhibit 
the phenomenon of spin-selective tunneling~\cite{AmashaSDT,Stano,Paaske,Katsaros} in a magnetic field. 

\bibitem{note1}
In the leads, the zero of energy is taken at the Fermi surface at equilibrium (hence $\mu_L=\mu_R=0$ at $eV=0$).

\bibitem{Trif}
M.~Trif, V.N.~Golovach, and D.~Loss,
Phys.\ Rev.\ B \textbf{77}, 045434 (2008).

\bibitem{Fasth}
C.~Fasth, A.~Fuhrer, L.~Samuelson, V.N.~Golovach, and D.~Loss,
Phys.\ Rev.\ Lett.\ \textbf{98}, 266801 (2007).

\bibitem{Nadj}
S.~Nadj-Perge, V.S.~Pribiag, J.W.G.~van den Berg, K.~Zuo, S.R.~Plissard, E.P.A.M.~Bakkers, S.M.~Frolov, and L.P.~Kouwenhoven,
Phys.\ Rev.\ Lett.\ \textbf{108}, 166801 (2012).

\bibitem{note2}
Strictly speaking, the spin-phonon coupling depends 
on the angle of the emitted phonon~\cite{Golovach}, and therefore, 
Eq.~(\ref{eqratewDnUp1pN}) is valid up to a factor stemming from that angular dependence.
However, it is exact in 1D.

\bibitem{Blum}
K. Blum, {\it Density Matrix Theory and Applications} (Plenum, New York, 1996), Chap. 8. 

\bibitem{Engel}
H.-A. Engel and D. Loss, Phys. Rev. Lett. {\bf 86}, 4648 (2001). 

\bibitem{RWA}
Equations~(\ref{system}) were found in the rotating wave approximation. 
The non-diagonal components of the density matrix were expressed through the diagonal ones
$$
\rho_{\Uparrow\Downarrow}=\frac{iM_{\Downarrow\Uparrow}^*a e^{-i\omega t}}{\Gamma+i(\Delta_Z-\omega)}(\rho_\Uparrow -\rho_\Downarrow).
$$
This solution is valid under the condition $\Delta_Z \gg \Gamma\gg M_{\Downarrow\Uparrow}\sqrt{N_Q}$.

\bibitem{mirrors}
We consider here the best-case scenario when the mirrors are ideal and perfectly reflecting, and estimate only the phonon decay time due to  intrinsic mechanisms.  

\bibitem{Kent1}
A. J. Kent, N. M. Stanton, L. J. Challis, and M. Henini, Appl.\ Phys.\ Lett.\  \textbf{81},  3497 (2002). 

\bibitem{Zeitler}
U. Zeitler et al., Phys.\ Rev.\ Lett.\ {\bf 82}, 5333  (1999). 

\bibitem{Kittel}
H. Kittel et al., Z. Phys. B - Condensed Matter {\bf 77}, 79 (1989). 


\end{thebibliography}
\end{document}